\documentstyle[twoside,fleqn,espcrc2,epsfig]{article}

\newcommand{\AmS}{{\protect\the\textfont2
  A\kern-.1667em\lower.5ex\hbox{M}\kern-.125emS}}

\newcommand{\be}{\begin{equation}}
\newcommand{\ee}{\end{equation}}
\newcommand{\bea}{\begin{eqnarray}}
\newcommand{\eea}{\end{eqnarray}}

\def\gtap{\;\raisebox{-.5ex}{\rlap{$\sim$}} \raisebox{.5ex}{$>$}\;}

\newcommand{\ml}{\overline{m_l}(2GeV)}
\newcommand{\ms}{\overline{m_s}(2GeV)}
\title{Light Quark Mass Determinations from the Lattice.}

\author{C.R. Allton\address{Department of Physics,
                            University of Wales, Swansea,
                            Singleton Park,
                            Swansea SA2 8PP,
                            U.K.}
\thanks{Preprint SWAT/161}}

\begin{document}

\begin{abstract}
This paper is a review of recent lattice determinations of the light
quark masses.
It describes the method employed to calculate quark masses in the
lattice formulation, and the extrapolations required to reach the
physical regime.
This review is designed to be accessible to a general audience, not
specifically lattice theorists.
\end{abstract}

\maketitle

\section{INTRODUCTION}

The masses of the light quarks are fundamental parameters of the
Standard Model, and yet are still surprisingly poorly determined. The
Particle Data Book gives their values as \cite{pdb},
\bea \nonumber
{\overline{m_u}(1GeV)} &=&   2 \rightarrow   8 \; MeV   \\ \label{eq:pdb}
{\overline{m_d}(1GeV)} &=&   5 \rightarrow  15 \; MeV  \\ \nonumber
{\overline{m_s}(1GeV)} &=& 100 \rightarrow 300 \; MeV
\eea

Over the last few years, several lattice projects have calculated these
quark masses \cite{ape_mass,lanl,fnal,ape_leon,sesam}. However, these
predictions have been mutually consistent only within a factor of two.
It is therefore vital that the reason for this uncertainty is
understood. This review describes the main reasons for the spread of
values in terms of (i) the parameters used by the different
collaborations, and, (ii) the extrapolations required to reach the
physical regime from the region simulated. It is very
reasonable to assume that the spread in lattice values can be greatly
reduced in the next couple of years once more precise calculations are
made, and the issue of the extrapolations is resolved. This would lead
to extremely accurate lattice determinations of the light quark masses,
far more constraining than eq.(\ref{eq:pdb}) above.

The bulk of lattice calculations have studied the isospin averaged
quark mass $m_l=\frac{1}{2}(m_u+m_d)$ and I will restrict my attention
to these calculations. I note though that there has been a preliminary
lattice investigation of the isospin mass splitting $m_d-m_u$
\cite{eichten} by using electromagnetic as well as gluonic fields. They
obtain a value for $m_u/m_d$ of $\approx 0.5$ in excellent agreement with other
methods showing that isospin splitting effects are also technically
accessible on the lattice.

In the next section I review the principles involved in lattice
calculations of light quark masses. I then discuss
the various extrapolations that are required in order to make contact
with the real world in section 3. This section will be the main focus of the review.
In section 4, I review the recent lattice results, attempting to
explain some of the inconsistencies that occur. In this section I give
rough lattice values of the light quark masses, based on the data
and the issues raised in section 3. In the last section I
compare these values with estimates using other methods.

For other reviews on this topic, including summaries of lattice results
prior to 1994 see \cite{paul_lat96,reviews}. I note also that
\cite{vittorio} and \cite{rajan_lat97} contain very recent reviews of
this topic.

\section{HOW THE LATTICE CALCULATES THE LIGHT QUARK MASSES}
\label{lattice}

The generic lattice approach to spectrum calculations studies
two-point correlation functions, $G_2(t)$, of hadronic interpolating
operators in a background sea of glue (and sea quarks in the case of
``unquenched'' calculations). These background configurations are weighted
by the appropriate Boltzmann factor (which means that the approach is
identical to calculation of correlation functions in statistical
mechanics). Lattice calculations are fully non-perturbative, and, in
principle, only involve approximations that can be systematically
improved. In practice the inclusion of dynamical sea quarks (i.e.
virtual quark - anti-quark pairs) is a technical headache because it
involves a huge increase in computational requirements. However, more
and more lattice results are being published with dynamical sea quarks
(also called ``unquenching'') and this trend is certain to gain
momentum.

In the calculation of $G_2(t)$, if one were fortunate enough to have
The Interpolating Operator of the hadron in question, then $G_2(t)$
would be The Two-point Function. However, when using a non-perfect
operator, ${\cal O}$, $G_2(t)$ receives contributions from all hadronic
states which have non-zero overlap with ${\cal O}$.
It is straightforward to show that $G_2(t)$ has the following form

\[
G_2(t) = \sum_{i} Z_i e^{-M_i t}
\]
where the sum is over the hadronic states, and $M_i$ and $Z_i$ are the
hadronic mass and overlap of the $i-$th state. Note that since the
calculation is performed in Euclidean space-time, the excited
states are exponentially suppressed with respect to the fundamental
state (i.e. the exponentials have real arguments).
This means that $G_2(t)$ becomes The Two-point Function only for
sufficiently large values of $t$. Obviously the parameters of the
ground state, in particular the mass $M_0=M$, can be extracted by fitting
$G_2(t)$ for $t$ sufficiently large, i.e.

\[
G_2(t) \rightarrow Z e^{-M t}
\mbox{\hskip 5mm as \hskip 5mm $t \rightarrow \infty$}.
\]
Here comes the crucial point. This mass, $M$, is not the mass of the
physical hadron. It is the mass of a hadron in a universe where the
quark masses are the same as those input into the lattice
calculation, i.e. the lattice is simulating a form of QCD, but not
the QCD of the real world.

In fact it is a little more complicated than this. The mass $M$ (and all
lattice predictions) are functions of {\em all} the input parameters of the
lattice calculation. Specifically these parameters are:

\begin{itemize}
\item quark masses, $m$;
\item lattice volume, $V$;
\item lattice spacing, $a$;
\item treatment~of~sea~quark~effects (i.e.~quenched~or~unquenched);
\item type of lattice action used.
\end{itemize}

In this section I consider only the dependence on the first
quantity above; I leave the other effects to the next section.

Obviously we do not know the correct values of the quark masses to
use in the calculation - these are, of course, the parameters that we
wish to calculate. So the next step is to {\em adjust} these light
quarks masses in the Lagrangian until the hadronic mass, $M$, agrees
with the experimental values. Thus, for instance, we equate:

\be
M_V(m_s,m_l) = M_{K^\ast} = 770 MeV,
\label{eq:mks}
\ee
where $M_V(m_s,m_l)$ is the lattice value of the vector meson composed
of quark with masses $m_s$ and $m_l$. We are
naturally free to use any hadron containing a strange quark in order
to determine the strange quark mass, e.g.\footnote{Normally baryons are
not used to set quark masses since their channels are more noisy in
lattice calculations.}
\bea \nonumber
M_V(m_s,m_s) &= M_{\phi} =& 1020 MeV, \\ \nonumber
\mbox{or} && \\ \label{eq:mk}
M_{PS}(m_s,m_l) &= M_K =& 494 MeV,
\eea
where $PS$ stands for pseudoscalar.
The differences in these various determinations of $m_s$ are
systematic effects and should be included in the error determination.
Note also that the determination of $m_s$ from the $K^{\ast}$ and
$\phi$ are generally in excellent agreement. This is due to the facts
that (i) the experimental values of these hadronic masses are linear in
their consituent quark masses, and
(ii) to a very good approximation, the lattice values of
$M_V(m_1,m_2)$ are also linear in $m_1$ and $m_2$.

Normally, due to the computational expense, the lattice calculation is
performed using only a small number (say 3 - 5) of quark masses. The
adjustment of the quark masses mentioned above is done by means of an
interpolation/extrapolation of these data points.

Note that there is an alternative approach to the calculation of light quark
masses on the lattice using the axial Ward identity (see \cite{ape_mass} \&
\cite{leonardo_lat97}). There has also been a very recent lattice recent calculation of the
``renormalisation group invariant mass'' in \cite{lueshcer_lat97}.

Once the values of the lattice quark masses have been found,
they are then expressed in the $\overline{MS}$ scheme, traditionally at
the scale $2\;GeV$. (For convenience, I note the
conversion factor between $2\;GeV$ \& $1\;GeV$ is
$\overline{m}(1\,GeV) / \overline{m}(2\,GeV) \approx 1.4$.)
The matching procedure uses the
relationship between the two schemes,

\be
\overline{m}(\mu) = Z_m(\mu a) m(a),
\ee
where $\overline{m}(\mu)$ is the quark mass in the $\overline{MS}$
scheme at the scale $\mu$.
$Z_m$ has been calculated in perturbation
theory \cite{zm} and its renormalisation group-improved form at
next-to-leading order has been given in \cite{ape_mass}.\footnote{
Recently the lattice component of $Z_m$ has been calculated
non-perturbatively \cite{leonardo_lat97} which means that the only
perturbative component remaining in $Z_m$ is in the continuum matching.}
Note that the expansion parameter $g^2$ relevant for the matching coefficient
$Z_m$ is a short-distance (i.e. small) coupling constant, so
perturbation theory should be fairly well behaved for this
quantity (barring ``tadpole'' contributions \cite{ape_mass}).
This contrasts with the large distance scales governing hadronic
quantities like $M$ (which require a non-perturbative tool
such as the lattice to calculate correctly).

There are two families of the lattice actions currently used, Wilson
and Staggered. While the staggered action has features that are
particularly endearing for chiral physics, it has the big disadvantage
that its ${\cal O}(g^2)$ coefficient in $Z_m$ is very large; typically
this term is 50 - 100\% of the zero order term \cite{paul_lat96}.
Therefore connecting any lattice quark mass value with continuum
schemes introduces unknown higher order effects which are uncontrolled.
For this reason I do not consider results from the staggered approach in
this review.

In the Wilson action, the quark mass is additively renormalised. The
connection between the quark mass parameter in the lattice Lagrangian,
$K^{-1}$, and the quark mass is

\be
m(a) = a^{-1} \; \frac{1}{2} ( \frac{1}{K} - \frac{1}{K_{c}} ),
\label{eq:mlat}
\ee
where $K_{c}$ corresponds to a massless pion, i.e.

\[
M_{PS}^2(K_{c},K_{c}) \equiv 0.
\]

All fields and couplings in the lattice action have had their
dimensions removed by multiplication of appropriate powers of $a$.
These powers of $a$ must be put back in all dimensionful predictions
(see eq.(\ref{eq:mlat})). The lattice spacing $a$ can be obtained by
fixing the lattice prediction of any dimensionful physical quantity
with its experimental value. Thus one of the physical `predictions'
from the lattice is used to set $a$ (this corresponds to setting the
scale $\Lambda_{QCD}$), and one lattice `prediction' is used for each
quark mass determination (see above). The spacing $a$ can be set from
any dimensionful physical quantity (that has been measured by
experiment!). Different physical quantities can lead to slightly
different values of $a$, and this systematic effect should again be
reflected in the error estimate.

In this section I have described, in principle, how lattice
calculations of light quark masses are performed. In the next section,
I review the details of this approach, in particular the effect that
the extrapolations in the quark masses, volume, lattice spacing etc.
have on the final lattice prediction.

\section{EXTRAPOLATIONS}

\subsection{Basic Ideas}
\label{basic}

In the previous section I stressed that there are several input
parameters in lattice calculations of strong interation physics: the
quark masses, volume, lattice spacing, type of lattice action used and
the number of quark flavours included in the sea, $N_f$.\footnote{The
quenched approximation corresponds to $N_f=0$.} Assuming that all of
these parameters were set to their physical value, then the lattice
would give absolute predictions of QCD. However, for technical reasons,
these input parameters are almost never set equal to their physical
values, and so the following extrapolations are required (note that in
lattice calculations, one is free to use sea quark masses, $m^{sea}$,
different to the ``valence'' quark masses, $m^V$):
\bea \nonumber
\{m^{sea}\} &\rightarrow& \{m_u, m_d, ... \} \\ \nonumber
\{m^V\} &\rightarrow& \{m_u, m_d, m_s, ... \} \\ \label{eq:extrap}
\mbox{Volume} &\rightarrow& \infty \\ \nonumber
a &\rightarrow& 0 \\ \nonumber
N_f &\rightarrow& \mbox{``}2\mbox{''}.
\eea

Thus the lattice simulates QCD, but with an unusual (i.e. unphysical)
set of parameters - the light quark masses are larger than their
physical values (typically they are $\gtap 50$ MeV), the volume is
finite, space-time is discretised, and the number of dynamical quark
flavours does not always correspond with the real world. It is
essential that all of the extrapolations eq.(\ref{eq:extrap}) are under
control in order for the lattice to make accurate determinations of
physical quantities.

A reliable extrapolation requires data points over a range of values of
the parameter in question. Thus, for example, an extrapolation in $a$
requires data at several values of $a$. Due to the present limitations
in computing power, it is not possible to cover a large amount of the
(multi-dimensional) parameter space in order to perform all of the
extrapolations. Therefore most lattice projects concentrate on the
effects of varying only one or two of the above parameters. Different
groups often concentrate their efforts in studying and extrapolating
different parameters and this fact can sometimes lead to different
final predictions for the quark mass. The wide variation in lattice
predictions for the quark masses referred to in the introduction is
primarily due to this effect.\footnote{In my opinion this point has not
been fully communicated to the non-lattice community.} Data {\em at the
same parameter values} from different groups are generally compatible.

Unfortunately, all of the extrapolations in eq.(\ref{eq:extrap}) are
required in order to  obtain The Lattice Prediction of the quark mass
(or, in fact, of any physical quantity). Thus, while the lattice
calculations performed so far are very encouraging, more work will be
needed to reach this ultimate goal. Fortunately, many physical
quantities aren't nearly as sensitive to these input parameters as the
quark mass, and therefore don't suffer from these extrapolation
effects to the same extent.

In the following subsections I will discuss the effects of
extrapolations in the lattice spacing, $a$, the sea quark mass,
$m^{sea}$, as well as the related systematics due to unquenching
effects. I do not consider volume effects, since they are relatively
small. The extrapolations in $m^V$ are basically those discussed in the
previous section.

\begin{figure}
\begin{center}  
\protect\label{fig:params}
\mbox{\epsfig{file=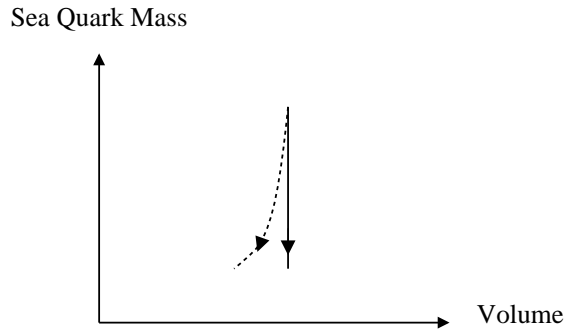, angle=270, width=7cm}}
\vskip 10mm
\caption{Schematic plot of the volume effect when naively extrapolating in
sea quark mass in unquenched simulations.}
\end{center}
\end{figure}

\subsection{Lattice Spacing Effects}
\label{cont_limit}

In \cite{lanl}, a collection of world data for $\ml$ and $\ms$ was
extrapolated to the $a=0$ limit. They found that the quenched Wilson
data had a strong dependence on $a$. Assuming a linear dependence
$\overline{m}(a) \sim \overline{m}(0) \times ( 1 + xa )$, the continuum
extrapolant, $\overline{m}(0)$, was as much as 40\% lower than the
simulated data. (This applied to both $\overline{m}_l$ and
$\overline{m}_s$.) An analysis of unquenched Wilson and staggered data
was presented as ``preliminary'' since these simulations were in their
early stages \cite{lanl}.

In \cite{fnal}, some evidence for an ${\cal O}(a)$ effect was found and
this was included their final quark mass estimate.

Another group, \cite{ape_leon}, analysed quenched Wilson and improved
Wilson $m_s$ data from the APE collaboration over a range of fairly
fine lattice spacing. They found that any ${\cal O}(a)$ effects were at
least as small as the statistical errors in the calculation and
therefore no extrapolation in $a$ was performed.

It is my belief that more data of better quality is required
to unambiguously determine the quark mass dependency on $a$.

\subsection{Sea Quark and Unquenching Effects}
\label{sea_quark}

A very interesting study of the dependency of $m_l$ and $m_s$ on the
sea quark mass was presented in \cite{sesam}. The calculation was at a
fixed value of the strong coupling constant (naively corresponding to a
fixed value of $a$) and used three values of the sea quark mass, the
heaviest being at around the strange quark mass, and the lightest being
approximately half $m_s$. A parallel calculation was performed at a
similar value of $a$ in the quenched approximation (i.e. corresponding
to $N_f=0$, or equivalently, $m^{sea} = \infty$). The dependence of
physical quantities such as meson masses on $m^{sea}$ was found to be
strikingly large. This implies that the lattice spacing obtained, for
instance, from $M_\rho$ (see Sect. \ref{lattice}) is itself a function of
$m^{sea}$.\footnote{The strong dependence on $m^{sea}$ on the static
quark potential and hadron masses has also recently been observed by
the UKQCD collaboration \cite{ukqcd_dyn}.}

A little thought shows that this dependence of $a$ on $m^{sea}$ should
not be surprising. It has long been known that, in order to maintain
roughly fixed physics (i.e. constant $a$ values) when comparing
quenched and unquenched data, the coupling $\beta=6/g^2$ had to be
renormalised by, say, $\sim 0.6$ (see e.g. \cite{degrand_lat90}). This
implies that, on the $m^{sea} - \beta$ plane, there is a curve of
constant $a$ connecting the quenched ($m^{sea}=\infty$) simulation at,
say, $\beta=6.0$ with the unquenched point (corresponding to a small
value of $m^{sea}$) with $\beta \approx 6.0 - 0.6$. Now consider
varying $m^{sea}$ while keeping $\beta$ fixed. This is the parameter
space investigated by the SESAM collaboration \cite{sesam}. This sweeps
along a line with a component {\em orthogonal} to the curve of constant
$a$. Therefore it is entirely reasonable that one finds a variation of
$a$ along this direction as reported by \cite{sesam}.

This feature of a variation of $a$ with $m^{sea}$ has interesting
implications. It means that as one extrapolates in $m^{sea}$ to the
physical value, $m_l$, the lattice volume {\em decreases} in physical
units (since $a$ decreases). This is displayed schematically as the
dashed curve in Fig.1. Normally one uses PCAC or quark model
considerations to model the behaviour of masses with $m_l$ - at {\em
fixed} volume which would correspond to solid vertical line in this
figure. In principle, this procedure will have to be modified to
compensate for the finite volume effects which mix with the $m^{sea}$
effects. However, in this relatively early stage of unquenched
calculations, one is forced to ignore such effects in order to make
progress.

The SESAM collaboration also has data at non-degenerate ``valence''
quark masses $m_1^V, m_2^V$ \cite{sesam}. This means that they are able
to apply eqs.(\ref{eq:mks} \& \ref{eq:mk}) directly to obtain $m_s$
rather than relying on SU(3) flavour symmetry (see, e.g.
\cite{ukqcd_nondeg}). 
Importantly also they have data for $m^V \ne m^{sea}$
and they present arguments about the correct procedure for extrapolating in
$m^V$ and $m^{sea}$ pointing out the perils of following an
incorrect procedure \cite{sesam}.
I reproduce their arguments here.

Their explanation can be understood by studying Fig.2 which is
taken from Fig.4 in the firsr reference in \cite{sesam}
(and slightly generalised). In this
plot, the sea and valence lattice quark mass parameters $1/K^{V}$ \&
$1/K^{sea}$ (see eq.(\ref{eq:mlat})) are plotted against each other. The
solid diagonal line corresponding to $K^{V}=K^{sea}$ corresponds to the
degenerate case where much of the earlier simulations have been
performed. The three diagonal dashed lines correspond to the
constraints $M_{PS}(K^{V},K^{sea}) = 0$, $M_{V}(K^{V},K^{sea}) =
M_\rho$, and $M_{V}(K^{V},K^{sea}) = M_\phi$ (labelled \fbox{1},
\fbox{2} \& \fbox{3} respectively).
Note that, as constructed in \cite{sesam}, these lines are
parallel, and empirically their slopes are $\approx -1$.
The points labelled ``$\rho$'' and
``$\phi$'' are the physical points; $\rho$ consisting of two light
quarks in a sea of light quarks, and the $\phi$ consisting of two
strange quarks in a sea of light quarks.
The point marked ``0'' corresponds to $m^V=m^{sea}=0$.

The two displacements on the
$y$-axis denote the {\em correctly determined} quark masses $m_l$ and
$m_s$ (see eq.(\ref{eq:mlat})) - i.e. they are calculated in a sea
consisting of physical ``$l$'' quarks.
The vertical displacements in the body of the figure show the {\em
incorrectly determined} values of $m_l$ and $m_s$.
The incorrect $m_l$ value, marked ``$m_l^Q$'', is obtained if one performed
a ``quenched-like'' analysis - i.e. an extrapolation in $m^V$ at {\em
fixed} $m^{sea}$.
Note that this is around twice the correct value of $m_l$.
The $m_s^Q$ value is the corresponding ``quenched-like'' value of $m_s$.
Note that this is surprisingly close to the correct $m_s$ value !
Finally, $m_s^{deg}$ is the $m_s$ value obtained if one worked
with {\em degenerate} sea and valence quarks.
This can be seen to be roughly half of the correct $m_s$ value.

\begin{figure}
\begin{center}  
\protect\label{fig:sesam}
\mbox{\epsfig{file=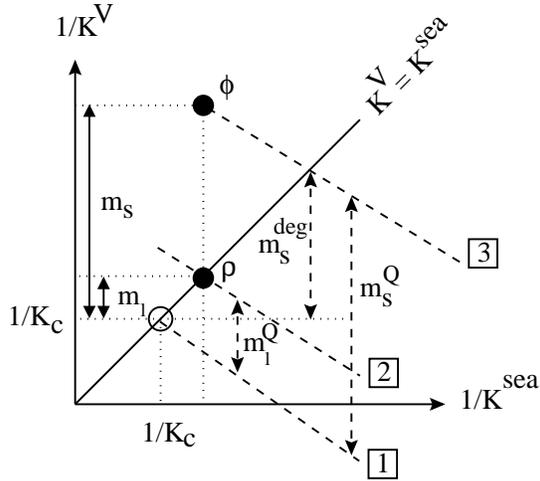, angle=270, width=7cm}}
\vskip 10mm
\caption{
A slightly generalised version of Fig.4 in
[6]
in which the sea and valence lattice quark mass parameters $1/K^{V}$ \&
$1/K^{sea}$ (see eq.(\ref{eq:mlat})) are plotted.
The three diagonal dashed lines correspond to
the constraints $M_{PS}(K^{V},K^{sea}) = 0$ (denoted \fbox{1}),
$M_{V}(K^{V},K^{sea}) = M_\rho$, (\fbox{2}), and
$M_{V}(K^{V},K^{sea}) = M_\phi$, (\fbox{3}).
The physical points are labelled $\rho$ and $\phi$ with the
corresponding correct values of $m_l$ and $m_s$.
``O'' corresponds to $m^V=m^{sea}=0$.
See Sect.\ref{sea_quark} for details.
}
\end{center}
\end{figure}

\subsection{Summary of Extrapolation Effects}

In concluding this section, it is clear that further work is required to
fully understand and compensate for the effects of the extrapolations
of the input parameters to their physical values in a lattice simulation.
As detailed above, both the continuum extrapolation
($a \rightarrow 0$, see Sect. \ref{cont_limit}) and the sea quark effects
(Sect. \ref{sea_quark}) introduce potentially large systematics that can be
as much as a factor of two. It is these extrapolations that are the main
source of the large spread of lattice values for the light quark masses
presented in contemporary literature.

Despite these uncertainty, lattice calculations performed by different
groups at common parameter values are in generally excellent agreement
leading to great optimism for the future.

\section{RECENT LATTICE RESULTS}

\begin{table*}[hbt]
\setlength{\tabcolsep}{.5pc}
\caption{ \it{Recent lattice results for the light quark masses.}
\label{tab:lat}}
\begin{tabular*}{\textwidth}{@{}l@{\extracolsep{\fill}}lllccccc}
\hline
\multicolumn{2}{l}{\bf Group}& $\ml$ & $\ms$ & Unquenched ? &
                Correct $m^{sea}$ & $a \rightarrow 0$ & ``Improved Action'' \\
          &                  & [MeV] & [MeV] &              &
                extrapolation ? & taken ?                     & used ?              \\
\hline
APE & \cite{ape_mass}	& 	& 128(18)	& No	& -	& - & Yes \\
APE & \cite{ape_leon}	&	& 122(20)	& No	& -     & ``Yes''& Yes \\
LANL & \cite{lanl}	& 3.4(4)(3)&100(21)(10)	& No	& -     & Yes	& Yes\& No \\
	&		&``2.7(3)(3)''&``68(12)(7)''&Yes& No    & Yes	&	\\
FNAL & \cite{fnal}	& 3.6(6)& 95(16)	& No	& -     & ``Yes''& Yes \\
	&		&``2.1$\rightarrow$3.5''&
				``54$\rightarrow$92''
						& Yes	& No    &	&	\\
SESAM & \cite{sesam}	& 5.5(5)& 166(15)	& No	& -     & No	& No \\
	&		& 2.7(2)& 140(20)	& Yes	& Yes   & No	& No \\
\hline
\end{tabular*}
\end{table*}

For the reasons outlined in Sect. \ref{lattice}, I will restrict myself
to discussing recent lattice results from the Wilson family of actions.
There have been many calculations of light quark masses over the years
(see \cite{paul_lat96,reviews} and references therein), but I will
concentrate on the more recent results.
These come from 4 groups
\cite{ape_mass,lanl,fnal,ape_leon,sesam} and are shown in Table \ref{tab:lat}.
(Note that the APE/Rome group has new data in \cite{leonardo_lat97}
which are compatible with those from \cite{ape_mass,ape_leon}.)
Alongside the values for $\ml$ and $\ms$ are indicated whether
(a) the simulations were performed in the quenched approximation,
(b) the correct $m^{sea}$ extrapolation was performed (see Sect.\ref{sea_quark})
(c) an attempt was made to allow for ${\cal O}(a)$ errors
(see Sect.\ref{cont_limit}), and,
(d) whether an ``improved action'' was used\footnote{These are lattice actions with
reduced discretisation systematics.}. Ideally, the answers to these
questions should all be ``Yes'', but, as can be seen, there is no one
simulation where this is the case. (See the discussion in Sect.
\ref{basic}.)

\begin{table}[h]
\caption{ \it{Recent non-lattice results for $m_l$ and $m_s$.}}
\label{tab:others}
\begin{center}
\begin{tabular}{ll}
\hline
\multicolumn{2}{c}{\bf QCD Sum Rules} \\
\hline
\cite{bpd}	& $\ml = 4.9(1.) MeV$ \\
\cite{dn}	& $3.4 MeV \le \ml \le 5.6 MeV$ \\
\cite{dn}	& $\ml \approx 5 MeV$ \\
\hline
\cite{jm}	& $\ms = 114(21) MeV$ \\
\cite{sn}	& $\ms = 141(21) MeV$ \\
\cite{cps}	& $\ms = 146(14) MeV$ \\
\cite{cfnp}	& $\ms = 102(13) MeV$ \\
\hline
\multicolumn{2}{c}{\bf Chiral Perturbation Theory} \\
\hline
\cite{leut}	& $m_s/m_l = 24.4(1.5) MeV$ \\
\hline
\multicolumn{2}{c}{\bf Experiment} \\
\hline
\cite{chen}	& $\ms \approx 160 MeV$ \\
\hline
\end{tabular}
\end{center}
\end{table}

The quotation marks around the unquenched results for the LANL and FNAL
groups signify that these are considered as preliminary \cite{lanl}, or
are based on early unquenched results and plausibility arguments
\cite{fnal}.

It is interesting to note that the small value for the unquenched $\ms$
in \cite{lanl} and \cite{fnal} are not reproduced in the fuller
unquenched analysis of \cite{sesam}. This is due to two extrapolation
systematics: (i) the sea quark effects are correctly taken account of
in \cite{sesam}, but not in \cite{lanl} (see the discussion surrounding
``$m_s^{deg}$'' in Sect. \ref{sea_quark}); and (ii) the
${\cal O}(a)$ effects are treated differently -  \cite{sesam} does no
$a \rightarrow 0$ extrapolation since only one value of the lattice
spacing was used (furthermore an {\em un-}improved action was used), whereas both
\cite{lanl} and \cite{fnal} incorporate a linear $a \rightarrow 0$
extrapolation procedure (see Sect. \ref{cont_limit}).

The other interesting feature to note is that the {\em relative}
quenching effects in $\ms$ appear small, whereas they appear
significant in the case of $\ml$ \cite{sesam}. This can again be
understood in terms of the discussion in Sect. \ref{sea_quark}
surrounding Fig.2.

By considering the extrapolation effects discussed in Sect. 3, I give
the following rough estimates for the values of the (unquenched !)
quark masses based on the results in Table \ref{tab:lat}.
\bea
\ml &=&  2 \rightarrow   5 \; GeV, \label{eq:mql} \\
\ms &=& 80 \rightarrow 140 \; GeV. \label{eq:mqs}
\eea
The $\ml$ range is chosed so that it covers the values in Table \ref{tab:lat}.
The $\ms$ estimate is obtained by considering the fact
that the LANL (unquenched) predictions are likely to be too low due
to their sea quark treatment, whereas the SESAM value is possibly
too high due to ${\cal O}(a)$ effects.
The values in eqs.(\ref{eq:mql},\ref{eq:mqs}) are obviously only meant
as a guide, and until the parameter space of lattice calculations is
more fully explored, it is not possible to perform all the
extrapolations required to determine The Lattice Prediction of $\ml$
and $\ms$.

\section{COMPARISONS WITH OTHER METHODS}

I give in Table \ref{tab:others} some determinations of light quark
masses from non-lattice methods. (These have all been run to the scale
$2\;GeV$.) Note the preliminary result from the
Aleph Collaboration for $m_s$, \cite{chen}. The lattice estimates in
eqs.(\ref{eq:mql}\&\ref{eq:mqs}) are in fairly good agreement with these
results, although they seem generally on the low side. This point has
been investigated in \cite{laurent} where lower bounds on $m_l$ and
$m_s$ are derived from fairly fundamental principles. The lower limit for
$\ml$ in eq.(\ref{eq:mql}) is in some conflict with the bounds in \cite{laurent}.
(See also \cite{yndurain}.)


\section*{ACKNOWLEDGEMENTS}

I would like to thank fellow members of the APE and UKQCD
collaborations for many interesting discussions, especially Leonardo
Giusti, Vittorio Lubicz, Mauro Talevi, Tassos Vladikas and Hartmut Wittig.
I also acknowledge very useful conversations with Henning Hoeber and
Rajan Gupta.
Financial support from the Nuffield Foundation is acknowledged.



\section*{DISCUSSIONS}

\noindent
{\bf A.P. Contogouris,} University of Athens

\noindent
{\em Some time ago, P. Lepage of Cornell Univ. proposed a method for
lattice calculations and claimed that it could be used in a small
computer, even a laptop(!). Did you use his method? And if not, why?
More generally, would you like to comment about the present stage of
Lepage's approach?}

\noindent
{\bf C.R. Allton}

\noindent {\em No, the calculations that I have described do not use
Lepage's ``coarse-lattice'' method. This is a proposal that adds extra
terms to the lattice action in order to reduce the effects of
discretisation - i.e. it is a so-called ``improvement'' scheme. The
important feature of Lepage's approach is that the coefficients of
these additional terms are set by a re-arranged perturbation series
(contrasting with other improvement methods which use a fully
non-perturbative technique). Lepage's approach has had success in reducing
the rotational non-invariance of the static quark potential, and in
quantities heavily dependent on that potential (such as physics of the
heavy sector). However, progress in the light quark sector has been
limited so far and this is the reason why the method has not been used
for the determination of light quark masses.}

\noindent
{\bf K.F. Liu,} University of Kentucky

\noindent
{\em I want to add to what Chris has said. While the improved action program
has been working remarkably well for the pure gauge sector and heavy
quarks for NRQCD, its progress for light quarks is still limited at
this stage.}


\end{document}